\documentclass{llncs}

\begin{document}

\title{A nearly tight memory-redundancy trade-off for one-pass compression}
\author{Travis Gagie}
\institute{Dipartimento di Informatica\\
    Universit\`{a} del Piemonte Orientale}
\maketitle

Individuals store as much information these days as companies did a few years ago, companies store as much as governments did, and governments store a good deal more than many people think they should.  As much as we store, though, even more floods past every minute and disappears, and algorithms to sift through these massive data streams have become a hot topic in recent years.  If we consider streaming algorithms to be those that work online and in sublinear memory,\,\cite{Mut05} then the well-known LZ77~\cite{ZL77} compression algorithm becomes one when implemented with a sublinear sliding window.  It follows from Wyner and Ziv's analysis~\cite{WZ94} that LZ77 with a superconstant window is still universal with respect to finite-order Markov sources, and it is easy to see universality is impossible with constant memory.  Therefore, in some sense, for thirty years we have had an optimal algorithm for compressing data streams.  Nevertheless, common sense says having the window grow too slowly must result in poor compression and, indeed, it provably slows the compression ratio's convergence to the entropy.

In a previous paper~\cite{GM07} we showed how LZ77 can be implemented with a slightly sublinear window and the same upper bound on convergence as plain LZ77, but that this is impossible with a sublogarithmic window.  Plain LZ77 itself converges fairly slowly, however, so we eventually turned our attention to other algorithms.  Grossi, Gupta and Vitter~\cite{GGV??} recently proved a very strong upper bound for an algorithm based on the Burrows-Wheeler Transform:
\begin{theorem} \label{thm:GGV??}
Let $s$ be a string of length $n$ over an alphabet of constant size $\sigma$.  Using \(O (n)\) time we can always encode $s$ in \(n H_k (s) + O (\sigma^k \log n)\) bits for all integers \(k \geq 0\) simultaneously.
\end{theorem}
The $k$th-order empirical entropy \(H_k (s)\) of $s$ (see, e.g.,~\cite{Man01}) is its minimum self-information per character with respect to a $k$th-order Markov source.  Equivalently, it measures our expected uncertainty about a character of $s$ given a context of length $k$, as in the following experiment: $i$ is chosen uniformly at random between 1 and $n$; if \(i \leq k\), then we are told the $i$th character of $s$; otherwise, we are told the $k$ preceding characters and asked to guess the $i$th.  Therefore, by the Noiseless Coding Theorem~\cite{Sha48}, we need at least \(n H_k (s)\) bits to encode $s$ with an algorithm that uses only contexts of length at most $k$.

In this note we use two facts about empirical entropy: first, \(n H_k (s)\) is super\-additive, i.e., \(|s_1 s_2| H_k (s_1 s_2) \geq |s_1| H_k (s_1) + |s_2| H_k (s_2)\); second, if every occurrence in $s$ of each $k$-tuple is followed by the same distinct character (or the end of the string), then \(H_k (s) = 0\).  The first fact means if we break $s$ into $b$ blocks and encode each block separately with Grossi, Gupta and Vitter's algorithm, then the bound on the encoding's total length is \(n H_k (s) + O (b\,\sigma^k \log n)\) for all integers \(k \geq 0\) simultaneously.  The second fact means a de Bruijn sequence of order $k$ --- or a string consisting of any number of repetitions of all but the last \(k - 1\) characters of such a sequence --- has $k$th-order empirical entropy 0.  A $\sigma$-ary de Bruijn sequence of order $k$ contains every possible $k$-tuple exactly once and, thus, contains \(\sigma^k + k - 1\) characters, of which the first and last \(k - 1\) are the same.  Over sixty years ago de Bruijn~\cite{deB46} counted the number of binary de Bruijn sequences of order $k$, and five years ago Rosenfeld~\cite{Ros02} generalized his result:
\begin{theorem} \label{Ros02}
There are $(\sigma!)^{\sigma^{k - 1}}$ $\sigma$-ary de Bruijn sequences of order $k$.
\end{theorem}
With these two theorems we can easily prove a nearly tight trade-off between memory and redundancy:
\begin{theorem} \label{trade-off}
Let $s$ be a string of length $n$ over an alphabet of constant size $\sigma$ and let $c$ and $\epsilon$ be constants with \(1 \geq c \geq 0\) and \(\epsilon > 0\).  Using \(O (n)\) time, \(O (n^c)\) bits of memory and one pass we can always encode $s$ in \(n H_k (s) + O (\sigma^k n^{1 - c + \epsilon})\) bits for all integers \(k \geq 0\) simultaneously.  On the other hand, even with unlimited time, using \(O (n^c)\) bits of memory and one pass we cannot always encode $s$ in \(O (n H_k (s) + \sigma^k n^{1 - c - \epsilon})\) bits for, e.g., \(k = \lceil (c + \epsilon / 2) \log_\sigma n \rceil\).
\end{theorem}

\begin{proof}
We first prove the upper bound.  Suppose we are given $n$ in advance.  Let $A$ denote Grossi, Gupta and Vitter's algorithm.  Although $A$ itself is not one-pass, we use it as a subroutine in a one-pass algorithm as follows: we process $s$ in \(O (n^{1 - c + \epsilon / 2})\) blocks \(s_1, \ldots, s_b\), each of length \(O (n^{c - \epsilon / 2})\); we read each block $s_i$ into memory in turn, compute and output \(A (s_i)\), and erase $s_i$ from memory.  Since $A$ takes linear time and, thus, memory at most proportional to the input size times the word size, we compute \(A (s_1), \ldots, A (s_b)\) using \(O (n)\) time, \(O (n^c)\) bits of memory and one pass.  As we noted above, by superadditivity the whole encoding is at most \(n H_k (s) + O (\sigma^k n^{1 - c + \epsilon})\) bits for all integers \(k \geq 0\) simultaneously.  Now suppose we are not given $n$ in advance.  We work as before but we start with a constant estimate of $n$ and, each time we have read that many characters of $s$, double it.  This way, we increase the number of blocks by an \(O (\log n)\)-factor and the size of the largest block by an \(O (1)\)-factor, so our asymptotic bound on whole encoding's length does not change.

We now prove the lower bound.  Suppose \(k = \lceil (c + \epsilon / 2) \log_\sigma n \rceil\), $d$ consists of all but the last \(k - 1\) characters of a randomly chosen $\sigma$-ary de Bruijn sequence of order $k$, and $s$ consists of repetitions of $d$.  Then \(H_k (s) = 0\) and \(O (n H_k (s) + \sigma^k n^{1 - c - \epsilon}) = O (n^{1 - \epsilon / 2})\), but $d$'s expected Kolmogorov complexity is at least \(\log (\sigma!)^{\sigma^{k - 1}} = \Omega (\sigma^k) = \Omega (n^{c + \epsilon / 2})\), i.e., at least linear in $d$'s length and asymptotically greater than the memory we can use.  Notice we can reconstruct $d$ from the memory configurations when we start and finish reading a copy of $d$ in $s$ and the bits we output while reading that copy (if there were another string $d'$ that took us between those two memory configurations while outputting those bits, then we could substitute $d'$ for that copy of $d$ without changing the overall encoding).  Therefore, we output \(\Omega (\sigma^k)\) bits for each copy of $d$ in $s$, or \(\Omega (n)\) bits in total. \qed
\end{proof}

\bibliographystyle{plain}
\bibliography{trade-off}

\end{document}